# Approaching the Intrinsic Bandgap in Suspended High-Mobility Graphene Nanoribbons

Ming-Wei Lin[1,#], Cheng Ling[1,#], Luis A. Agapito[2,#], Nicholas Kioussis[2], Yiyang Zhang[1,3], Mark Ming-Cheng Cheng[3], Wei L. Wang[4], Efthimios Kaxiras[4] and Zhixian Zhou[1,*]

[1]Department of Physics and Astronomy, Wayne State University,
Detroit, MI 48201, USA

[2]Department of Physics, California State University, Northridge, CA 91330, USA

[3]Department of Electrical and Computer Engineering, Wayne State University,
Detroit, MI 48202, USA

[4]Department of Physics and School of Engineering and Applied Sciences,
Harvard University, Cambridge, Massachusetts 02138, USA

\# These authors contributed equally.

*Author to whom correspondence should be addressed, electronic mail: zxzhou@wayne.edu

ABSTRACT: We report electrical transport measurements on a suspended ultra-low-disorder graphene nanoribbon(GNR) with nearly atomically smooth edges that reveal a high mobility exceeding 3000 cm$^2$ V$^{-1}$ s$^{-1}$ and an intrinsic band gap. The experimentally derived bandgap is in *quantitative* agreement with the results of our electronic-structure calculations on chiral GNRs with comparable width taking into account the electron-electron interactions, indicating that the origin of the bandgap in non-armchair GNRs is partially due to the magnetic zigzag edges.

**Introduction**

Graphene is a single atomic layer of three-fold coordinated π-bonded carbon atoms that exhibits exceptionally high carrier-mobility, offering the tantalizing possibility of all-carbon electronics[1]. As an infinite two-dimensional solid, graphene is a zero-gap semiconductor with finite minimum conductivity, which poses a major problem for conventional digital logic applications. To overcome this bottleneck, many theoretical and experimental studies have focused on engineering an energy gap in graphene. A tunable band gap up to 250 meV can be induced by a perpendicular electric field in bilayer graphene[2]. A band gap can also be created by strain[3] or by chemical modification of graphene[4]. More generally, a band gap can be created by spatial confinement and edge effects[5]. Louie *et al.*[6] showed theoretically that grapheme nanoribbons (GNRs) with pure armchair or zigzag shaped edges always have a nonzero and direct bandgap, the value of which depends on the ribbon crystallographic orientation and edge structure. In lithographically patterned GNRs with varying widths and crystallographic orientations, electrical transport studies established the presence of a width-dependent transport gap[7, 8]. Several possible mechanisms have been proposed to explain the transport gap observed in GNR-based field-effect transistors(GNR-FETs), including re-normalized lateral confinement due to localized edge states[7, 8], percolation driven metal-insulator-transition caused by charged impurities[9], quasi-one-dimensional Anderson localization[10], and Coulomb blockade due to edge-roughness[11]. More recent experimental studies on disordered GNRs further indicate that charge transport in the conduction gap of GNRs is likely dominated by hopping through localized states[12] or isolated charge puddles acting as quantum dots[13]. A significant increase in mobility has been observed in high-quality GNRs with nearly atomically smooth edges partially due to reduced edge scattering[14]. However, a large discrepancy remains between the bandgap extracted from these high-quality GNRs and that observed in other reports[15], even though the GNRs were synthesized using a similar approach. This discrepancy may be attributed to different edge structures, but could also be due to extrinsic conduction through defects and impurity states within the bandgap[2, 16].

In this paper, we report the first variable-temperature electrical-transport study of suspended ultra-low-disorder GNRs with nearly atomically smooth edges. Suspension of the GNRs not only removes the substrate influence but also allows a thorough removal of impurities, including those trapped at the interface between the GNR and the substrate, leading to a substantial increase of the carrier mobility. We observe high mobility values exceeding 3000 cm$^2$ V$^{-1}$ s$^{-1}$ in GNRs that of width~20, the highest mobility value reported to date on GNRs of similar dimensions. Furthermore, we demonstrate that the activation gap extracted from the simple activation behavior of the minimum conductance and residual carrier density at the charge neutrality point approaches the intrinsic bandgap in ultra-low-disorder GNRs. In contrast to the results reported here, in typical transport measurements in GNRs the presence of non-negligible amount of disorder obscures the observation of the intrinsic bandgap. Moreover, the size of the bandgap derived from the transport measurements is in *quantitative* agreement with the results of our complementary tight-binding calculations for a wide range of chiral angles characterizing the GNR structure, supporting our proposed explanation, namely that the underlying electronic origin of bandgap enhancement is the magnetic nature of electronic states associated with zigzag edges.

**Experimental details**

The GNRs were produced by sonicating mildly-oxidized multiwall carbon nanotubes (MWNT) in a 1,2-dichloroethane (DCE) solution of poly(m-phenylenevinylene-co 2, 5-diy octocy- p-phenylenevinglene) (PmPV), where the PmPV is used as a surfactant to stabilize the unzipped GNRs in solution[14]. The solution was then centrifuged at 15000 rpm (Fisher Scientific Marathon 26kmr centrifuge) for 1 hr to remove aggregates and some of the remaining MWNTs, and a supernatant containing nanoribbons and remaining MWCTs was obtained. Next, the GNR samples from the supernatant were deposited on degenerately doped Si subtracts with 290 nm of thermal oxide. Non-contact mode AFM (Park System XE-70) measurements were used to locate individual GNRs with respect to the prefabricated Au alignment marks and to characterize their thickness, width and length. The GNRs produced from this method mostly consist of 1-3 layers. To determine the width, we have taken into account the AFM tip dilation effect (leading to artificial width increase) based on the estimated tip radius provided by the tip manufacturer.

FET devices consisting of individual GNRs are fabricated on Si substrates with 290 nm of thermal oxide using standard electron beam lithography and thermal deposition of 0.5 nm of Cr and 50 nm of Au, where the Si substrate is used as a back gate. Suspension of the GNRs in FET devices is achieved by placing a small drop of 1:6 buffered hydrofluoric acid (HF) on top of the GNR device for 90 s to etch way approximately 150 nm of the $SiO_2$ underneath the ribbons[17, 18]. The devices are annealed in vacuum at 600 °C for 10 minutes to clean the suspended ribbons and improve the electrical contacts before transferred to a Lakeshore Cryogenics vacuum probe station for further removing adsorbed impurities by current annealing and subsequent transport measurements in high vacuum (~$10^{-6}$ torr). The residual impurities on GNRs are gradually removed by repeatedly passing a large current through the ribbon; the final amount of impurities of the GNRs depends both on initial amount and the degree of current annealing.

A semiconductor parameter analyzer (Keithley 4200) was used to apply the annealing current and to measure the device characteristics for 4.3 < T < 300 K. We repeatedly applied gradually increasing annealing current and subsequently carry out the electrical measurements *in situ* after every consecutive step. The degenerately doped Si substrate was used as a back gate. To avoid possible collapsing of the suspended GNRs, the back-gate voltage $V_g$ was limited to the range – 15 V < $V_g$ < +15 V during the electrical measurements.

**Results and discussions**

We have fabricated over 20 suspended GNR-FET devices from GNRs synthesized by unzipping high quality multiwall carbon nanotubes[14]. A schematic diagram and an atomic force microscopy (AFM) image of a typical suspended GNR device are shown in the right and left insets of Fig. 1, respectively. As most of the devices were eventually damaged during the *in situ* currently annealing (likely caused by structural reconstruction at the defect sites), we report detailed electrical transport results on three high quality samples(samples A, C and D) characterized by extremely low disorder and compare them with those of a sample that contains a non-negligible amount of defects (sample B).

In Fig. 1we show the resistance ($R$) as a function of gate voltage ($V_g$) at different temperatures for two devices fabricated from a single uniform GNR. The GNR channels in these two devices have similar length (~600 nm), and nearly identical width (~20 nm) and thickness (1.4 nm corresponding to about 2 layers[14]) as determined by AFM before suspension[19]. Although both devices show characteristic ambipolar behavior arising from the electron-hole symmetry of graphene, they also exhibit remarkable differences. First, the resistance peaks at the charge neutrality point (CNP) in sample A are substantially sharper [Fig. 1(a)] than in sample B [Fig. 1(b)]. The full-width at half-maximum (FWHM) for sample A is more than an order of magnitude smaller than that for sample B at 160 K. Second, the maximum resistance at the CNP in sample A increases more rapidly with decreasing temperature than in sample B. These differences can be attributed to lower degree of disorder in sample A than in sample B. Defects, such as adsorbed charged impurities and structural imperfection, are expected to generate random potential fluctuations in the GNRs, which induce electron-hole puddles close to the CNP[17, 18]. As a result, the effect of gate voltage near the CNP is largely limited to the redistribution of charge carriers between electrons and holes without changing the overall carrier density. Therefore, a higher tunability of charge carriers near the CNP (and hence a much sharper resistance peak) is expected in samples with lower disorder. Similarly, the effect of thermally excited electron-hole pairs is also significantly enhanced with lower disorder, leading to a stronger temperature dependence of the maximum resistance.

We next focus on the influence of disorder on the carrier mobility and bandgap of GNRs. To extract accurate values for these quantities, we subtract the contact resistance from the total resistance using the following model to fit the $R(V_g)$ data:

$$R_{total} = R_{contact} + R_{channel} = R_{contact} + \frac{L/W}{ne\mu} \qquad (1)$$

Here, $R_{contact}$ and $R_{channel}$ are the metal/GNR contact resistance and GNR channel resistance, respectively[20]; $L$ and $W$ are the channel length and width, respectively; $\mu$ is the carrier mobility, and the carrier concentration $n$, can in turn be determined by the expression,

$$n = \sqrt{n_o^2 + C_g(V_g - V_{CNP})^2} \;, \qquad (2)$$

with $n_o$ being the residual carrier concentration at the maximum resistance, $C_g$ the back-gate capacitance (estimated to be $\sim 3\times10^{-8}$ F/cm$^2$ based on the capacitance of GNR-FET devices with similar ribbon width and taking into account the reduced dielectric constant due to the removal of ~150 nm of thermal oxide underneath the ribbon[21, 22]), and $V_{CNP}$ is the gate voltage at the charge neutrality point[20, 23]. As shown in Fig. 1, this model fits our experimental data reasonably well, especially in the hole-branch ($V_g < V_{CNP}$). The slightly lower conductance and minor deviation from the fitting at the electron side is likely due to the residual surface impurities and/or electrode metal doping[18, 21, 24]. From the fitting, a contact resistance of 30 ~ 70 kΩ is extracted, which is comparable to the value determined by 4-terminal measurements of similar GNRs devices (data not shown). Although this model assumes a gate-independent contact resistance, we believe this is a reasonable assumption for our devices given the nearly ohmic contact (except at low temperatures and near the CNP) and reasonably good fit of the data to the model, which is also consistent with the findings of Russo *et al.*[25].

Fig. 2 shows the mobility values derived from the fit as a function of temperature for samples A and B. The mobility of sample B has relatively weak temperature dependence and reaches ~ 1500 cm$^2$ V$^{-1}$ s$^{-1}$, in excellent agreement with that derived from substrate-supported GNRs synthesized using the same method[14]. Remarkably, the mobility of sample A increases from ~ 2000 cm$^2$ V$^{-1}$ s$^{-1}$ to over 3000 cm$^2$ V$^{-1}$ s$^{-1}$ as the temperature is lowered from 295 K to 150 K, suggesting that the mobility in this temperature range is largely limited by acoustic phonon scattering[26]. The peak mobility in sample A is the highest reported to date for GNRs of comparable widths[14], which can be attributed to the nearly atomically smooth edges and extremely low disorder. Below 150K, the mobility decreases with decreasing temperature, suggesting that the presence of a small amount of remaining disorder can play an increasingly important role at low carrier density (see detailed discussion below). Equally high mobility is also observed in sample C (data not shown). From the transfer characteristics, the field effect mobility of sample A in the hole region can be estimated as:

$$\mu = [\Delta G \times (L/W)]/(C_g \Delta V_g), \qquad (3)$$

where $G$ is the low-bias conductance of the sample[27] and the other parameters are defined in Eqs. (1) and (2). The field-effect hole mobility as a function of temperature for sample A is shown as "hollow squares" in Fig. 2, in reasonable agreement with the mobility values derived from the other method.

In an ideal intrinsic semiconductor without impurities, the conductance at the CNP, $G_{min}$ is expected to be dominated by thermally activated carriers and to vary with temperature as $G_{min} \propto \exp(-E_g/2k_BT)$, where $k_B$ is the Boltzmann constant and $E_g$ is the activation energy for electron excitation that corresponds to the bandgap. However, other mechanisms such as one-dimensional (1D) nearest neighbor hoping (NNH) through localized states in disordered GNRs may also lead to simple activated behavior of $G_{min}$[12]. To confirm that the activation energy derived from the temperature dependence of $G_{min}$ is indeed the intrinsic bandgap, it is necessary to show the same simple activation temperature dependence of the minimum carrier density ($n_0$) at the CNP (to first order approximation): $n_0 \propto \exp(-E_g/2k_BT)$. As shown in the Arrhenius plots in Fig. 3 (a) and (b), the $G_{min}$ and $n_0$ data from samples A and C (the latter being yet another low-disorder sample with $W \sim 37$ nm, $d \sim 2$ nm, and $L \sim 700$ nm) fit the simple activation model fairly well with a consistent activation energy gap of $E_g$ (A) = ~ 99meV(from $G_{min}$) and ~106 meV (from $n_0$) for sample A, and $E_g(C)$= ~ 55 meV (from $n_0$) and ~ 58 meV (from $G_{min}$) for sample C, respectively. Simple activation behavior is also observed in the residual carrier density of sample D ($W$ ~ 23 nm and $d$ ~ 1.6 nm), yielding a gap of 96 meV (data not shown). Furthermore, comparison of the $E_g$ values of samples A, C, and D demonstrates that the bandgap in our ultra-low-disorder samples is approximately inversely proportional to the ribbon width, consistent with theoretical predictions[6].These consistent results on multiple ultra-low-disorder GNR-FET devices strongly suggest that the intrinsic bandgap is approached.

On the other hand, $G_{min}$ and $n_0$ in sample B exhibit a much weaker temperature dependence than in samples A or C; forcing the simple activation law fit through the data of sample B yields a much smaller activation energy and corresponding bandgap of $E_g$~ 10 meV from both the $G_{min}$ and $n_0$ data. The large discrepancy between samples A and B is quite puzzling, since they are simply two different regions of the same GNR with highly uniform width and thickness and likely having the same nominal edge structure. The primary known difference between them is that sample A has lower disorder than sample B due to the spatial variation of disorder (such as remaining adsorbed impurities and structural defects which could be inherent in the original carbon nanotubes and/or introduced during the conversion from carbon nanotubes to GNRs). Given the small dimensions of the devices, even a small amount of disorder may play a significant role in their transport properties. Additionally, Au-contact doping may also vary from device to device. However, electrode doping is unlikely to be the dominant mechanism given that samples A and B not only have nominally identical contact structure and layout but also share a common electrode. Therefore, the weaker temperature dependence of $G_{min}$ and $n_0$ observed in sample B is likely to be due to extrinsic conduction through defects and carrier doping from charged impurities, similar to the bilayer graphene[2, 16]. An alternative explanation is that the presence of disorder weakens the on-site Coulomb interaction, which is largely responsible for the opening of a gap in the band structure of GNRs with zigzag edges[28]. Zigzag edges have indeed been observed by scanning tunneling microscopy(STM) in GNRs synthesized using the same method[29]; the smaller values of the bandgap found in these studies can be attributed to the reduced on-site Coulomb repulsion due to screening from the gold substrate[29]. It is also worth noting that the data for samples A, C (Fig. 3) and D (data not shown) start to deviate from the simple activation behavior below 100 K and the fit eventually breaks down below 77 K. The break down of the simple activated behavior at low temperatures can be attributed to extremely low residual carrier density: the value $n_0 \sim 7 \times 10^9$ cm$^{-2}$ at 77 K observed in sample A corresponds to only "one electron" in the device channel. Therefore, the residual carrier density (thus also the minimum conductivity) below 77 K is no longer determined by thermal activation.

In order to further verify that the simple activation gap observed in our ultraclean GNRs is the intrinsic bandgap (due to the extended states carrying current via thermal activation across the intrinsic bandgap), we compare the activation gap energy with the energy associated with the transport gap ($\Delta V_g$). The transport gap is correlated to an energy gap in the single particle spectrum given by:

$$\Delta_m = \frac{h}{2\pi} v_F \sqrt{\frac{2\pi C_g \Delta V_g}{e}}, \qquad (4)$$

where $v_F = 10^6$ m/s is the Fermi velocity of graphene and $C_g$ is the capacitive coupling of the GNR to the back gate. In disordered GNRs, where the electrical transport is dominated by the hopping between localized states, $\Delta_m$ is expected to be substantially larger than $E_g$[12]. In contrast, in highly ordered GNRs with very low impurity concentration $\Delta_m$ should be comparable to the intrinsic bandgap[15]. $\Delta V_g$ in this study is defined as the width of the back gate voltage region determined by a sudden increase of the slop in the $G(V_g)$ curve close to the CNP. As shown in Fig. 4(a), the $G(V_g)$ curve for sample A measured at 30 K yields a $\Delta V_g \sim 1.6$ V and hence $\Delta_m \sim 90$ meV, in reasonable agreement with the values of $E_g$ obtained from $G_{min}$ and $n_0$, indicating that the transport gap is associated with the large intrinsic bandgap. The linear dependence of $G$ on gate voltage $V_g$ at high temperatures [Fig. 4 (a), where the contact resistance is excluded] suggest that the field-effect mobility remains nearly constant as the carrier density changes and that the charge transport is limited by long-range scattering[30].

The transport gap can be alternatively probed by measuring the current-voltage (I-V) characteristics at various gate voltages. Fig. 4(b) shows representative I-V curves of sample A measured at 4.3 K. At gate voltages away from the CNP, the I-V curves are essentially linear. Near the CNP ($V_g$ = 1V), however, the I-V characteristic becomes strongly non-linear when the chemical potential of the GNR is within the transport gap. A nonlinear gap can be defined by the distances between two interception points made by fitting straight lines to both the low conductance region at low bias voltage and the high conductance region at high bias voltage, as shown in Fig. 4(b). The nonlinear gap (e$\Delta V_{ds}$) for sample A is approximately 60 meV, slightly smaller than the activation gap or the energy associated with the transport gap, which can be attributed to the fact that the gate voltage at which the nonlinear gap is measured slightly differs from the exact CNP. Unlike in highly disordered GNRs, where the presence of localized states and the formation of isolated charge puddles (which act as quantum dots) complicates the interpretation of the nonlinear gap in their I-V characteristics[12, 13], the nonlinear gap in our low-disorder GNRs may be approximated as the intrinsic bandgap for $V_g = V_{CNP}$[7].

In order to elucidate the underlying electronic origin of the high bandgap value in ultra-low-disorder GNRs, we carried out tight-binding (TB) calculations in model GNRs of comparable width (~20nm). Ultraclean GNRs with ultrasmooth edges are expected to be highly crystallographic and the measured intrinsic bandgap should be comparable to the theoretical values that assume periodicity. Because of the lack of information on the chirality($n, m$) of our ribbons, we calculated GNRs of a wide range of chiral angles($\theta$), varied from $\theta = 0°$ (zigzag GNR) to $\theta = 30°$ (armchair GNR) as shown in Fig. 5(a); GNRs with intermediate chirality exhibit mixed edges (zigzag/armchair) with dominant zigzag or armchair character as $\theta \to 0°$ or $\theta \to 30°$, respectively.

As seen in Fig. 5(a), the GNRs structures used in the calculations are derived from unzipping a CNT along the chiral unit-cell translational vector $\overrightarrow{OA} = (n, m)$ that determines the chiral angle $\theta$. The translational vector in turn restricts the width of the ribbons to discrete values that are the multiples of $|\overrightarrow{OB}|$, which is the minimum circumference of a $(n,m)$-type CNT. The electronic-structure calculations employ the single-band Hubbard model within the mean-field approximation:

$$\hat{H} = -t\sum_{\langle i,j\rangle,\sigma}\left(\hat{c}_{i\sigma}^{\dagger}\hat{c}_{j\sigma} + h.c.\right) + U\sum_i(\hat{n}_{i\uparrow}\langle\hat{n}_{i\downarrow}\rangle + \hat{n}_{i\downarrow}\langle\hat{n}_{i\uparrow}\rangle) \quad (5)$$

where $t$ is the hopping matrix element between nearest-neighbor sites $i$ an $j$, $\langle\hat{n}_{i\sigma}\rangle$ the expectation value of the number operator ( $\hat{n}_{i\sigma} = \hat{c}_{i\sigma}^{\dagger}\hat{c}_{i\sigma}$) on atom $i$ with spin $\sigma = \uparrow,\downarrow$, and $U$ is the on-site Coulomb interaction. The choice of the $t$ and $U$ parameters is crucial to making comparisons between experimental and theoretical values for the bandgap which is proportional to $(U/t)$. Furthermore, the values of $U$ and $t$ $t$depend on the choice of the exchange-correlation functional used in the density functional theory (DFT) calculations. We have used the *ab-initio* parameters ($t$= 3.2 eV and $U$ = 2$t$) reported by Pisani*et al.*[31], derived from fitting the antiferromagnetic band structure of GNRs using the fully-nonlocal functional (B3LYP) of DFT calculations, which includes a contribution of Fock exchange that compensates for the electronic self-interaction. Previous studies have shown that B3LYP is better suited than local, nonlocal or even other fully-nonlocal functionals to account for molecular magnetism[32]. These values are somewhat larger than those commonly employed in literature[33, 34], derived from DFT calculations employing local or nonlocal functionals. These values are also more appropriate to our suspended GNR samples that interact neither with a metallic substrate[29], which reduces $U$ through screening with the conduction electrons, nor with oxide substrates ($SiO_2$), which have much higher dielectric constant than air.

In the absence of electron-electron correlations ($U$ = 0 eV in Eq. (5)) the systems are non-magnetic. Interestingly, we find that the carbon atoms in the zizag chains in the mixed-edge

GNRs (even a single one per unit-cell in the limit when $\theta \rightarrow 30°$) introduce non-bonding states whose origin is topological frustration[35]. These non-bonding states form dispersionless "flat" bands at the Fermi level and render the systems gapless or metallic. For $U = 0$ eV, the band structure and density of states for all our systems follow the same pattern as for the $(n, m) = (4,2)$ GNR shown in Figs. 6(a) and 6(b) (red curves). This is analogous to the predicted presence of non-bonding states in randomly shaped 0-D graphene dots that contain combined zigzag/armchair edges[36]. Therefore, only "pure" armchair ribbons could sustain an energy bandgap which is not of magnetic origin. When electron-electron correlations are introduced ($U > 0$ eV) local magnetism arises along the edges of the ribbon, as seen in Fig. 5(b). Noticeably, the magnetization is predominantly higher on the zigzag sites than on armchair sites. Also, the flat bands split, opening an energy gap at the Fermi level (blue curves in Fig. 6).

Fig. 6(c) shows the energy bandgap and maximum spin magnetization for ~20 nm-wide GNRs with crystallographic orientations given by $\theta = $ 0°, 6.59°, 8.95°, 19.11°, 23.41°, 30°, corresponding to $(n, m) = $ (6,0), (7,1), (5,1), (4,2), (3,2), (3,3), respectively. Magnetic pure zigzag and non-magnetic pure armchair GNRs exhibit similar bandgaps of ~71 meV. Interestingly, for all the mixed-edge GNRs (with $0° < \theta < 30°$) the bandgap varies between 71 and 128 meV, in agreement with the experimentally determined bandgap for sample A, suggesting that the origin of the bandgap for mixed-edge GNRs is associated with the magnetism of the zigzag edges. The increase of bandgap in the zigzag-rich region ($\theta \sim 0°$) of Fig. 6(c) is consistent with an increasing insulating character caused by gradually breaking the zigzag π-network as the crystallographic orientation departs from $\theta = 0°$. As the chirality approaches the armchair-rich region ($\theta \sim 30°$) the spin magnetization quenches monotonically and the splitting induced by the second term of Eq. (5) becomes weaker, leading to a decreasing bandgap.

Our calculations were performed on single layer GNRs while the GNRs used in our experiment may consist of more than one layer. Nevertheless, the experimental and the theoretical bandgaps are still in good quantitative agreement. A likely scenario is that the experimentally derived bandgap is an average of the contributions from individual layers that have comparable bandgap values, which can be attributed to the combined effects of the relatively weak interlayer-interactions between non-AB (Bernal)-stacked layers[37] and the weak chirality dependence of the bandgap. Furthermore, moderate tensile strain may be present in our suspended GRNs as indicated by the lack of sagging (Fig. 1 inset), which is expected to slightly modify the size of the bandgap[38]. For the case of zigzag GNRs, a moderate strain leads to slight increase of the edge spin polarization, thus increasing the bandgap[38]. Therefore, the bandgap in our suspended chiral GNRs may be further enhanced by tensile strain.

**Summary**

In summary, we have fabricated GNRs with very low disorder by: (i) unzipping high quality CNTs with very low concentration of structural defects known to produce GNRs with nearly atomically smooth edges[14]; (ii) suspending the GNR from the substrate; and (iii) removing the remaining impurities by *in situ* current annealing. These ultraclean and ultra-smooth-edged GNRs not only exhibit high mobility exceeding 3000 $cm^2$ $V^{-1}$ $s^{-1}$, but also reveal the intrinsic electronic structure (bandgap) of GNRs. The good *quantitative* agreement between the experiment and theory suggests that the underlying mechanism responsible for the large bandgap in ultraclean suspended GNRs is most likely the magnetism associated with the zigzag edge components, which is strongly enhanced by the absence of either metallic or insulating substrates. The possible strain in the suspended GNRs may further augment the bandgap. Additional studies are underway to explore the tuning of the electronic and magnetic properties of such ultraclean GNRs via external electric and magnetic fields.

**Acknowledgement**

Z.Z acknowledges the support of the Wayne State University startup funds. Part of this research was conducted at the Center for Nanophase Materials Sciences under project # CNMS2009-044. The work at CSUN was supported by NSF-PREM grants DMR-00116566 and DMR-0958596. The work at Harvard was supported through the Nanoscale Science and Engineering Center, an NSF-funded facility. The authors also thank Boris Nadgorny for helpful discussion.

**Fig 1** Resistance versus gate voltage for: (a) sample A (lower-disorder), and (b) sample B, measured at various temperatures. The solid lines are the model fitting. The two samples belong to a single GNR with uniform width ($W \sim 20$ nm) and thickness ($d \sim 1.4$ nm), and both have the same length:$L \sim 600$ nm. Insets: schematic illustration of a GNR-FET consisting of a suspended GNR (right) and the contact electrodes, and AFM image of a GNR suspended by Au electrodes (left).

**Fig. 2** Mobility as a function of temperature for samples A and B. The solid squares and solid circles are the mobility extracted from the model fitting in Fig. 1; the hollow squares are the field effect mobility. The dashed lines are a guide to the eye.

**Fig. 3** (a) Temperature dependence of the minimum conductance for samples A, B and C. (b) Temperature dependence of the residual carrier density extracted from the model fitting in Fig. 1 for samples A, B and C. The solid lines are the fit to the simple activated behavior.

**Fig. 4** (a) Conductance versus gate voltage measured at various temperatures for sample A. (b) *I-V* characteristics measured at different gate voltages and at $T = 4.3$ K for sample A.

**Fig. 5** (a) Unrolled projection of a $(n,m)$-CNT of minimum circumference ($|\overrightarrow{OB}|$). The chiral angle $\theta$ is determined by the translational vector $\overrightarrow{OA} = (n,m) = (3,2) = 3\overrightarrow{a_1} + 2\overrightarrow{a_2}$. (b) Cross-section of a (3,2)-GNR with ~20nm width. The periodic unit-cell used in the calculation is shown shaded in green. The zoom-in regions show the spatial distribution of spin-up (cyan) and spin-down (red) magnetization. The magnitude of the magnetization is given by the radii, with the largest radius corresponding to spin magnetization $0.13\ \mu_B$.

**Fig. 6** (a) Electronic band structure, and (b) density of states of a (3,2)-GNR; results in the absence ($U = 0$ eV) and presence ($U = 6.4$ eV) of electron-electron correlation are shown in red and blue, respectively. (c) Calculated bandgaps (black) and maximum spin magnetization (red) for GNRs of $(n,m)$-type (6,0), (7,1), (5,1), (4,2), (3,2), and (3,3), corresponding to chiral angles $\theta$ in ascending order.

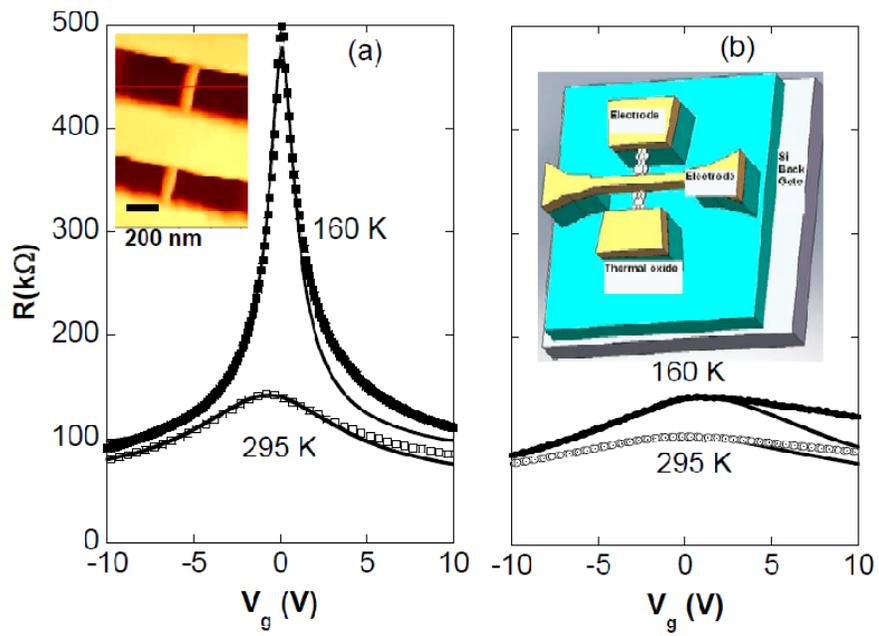

Fig. 1

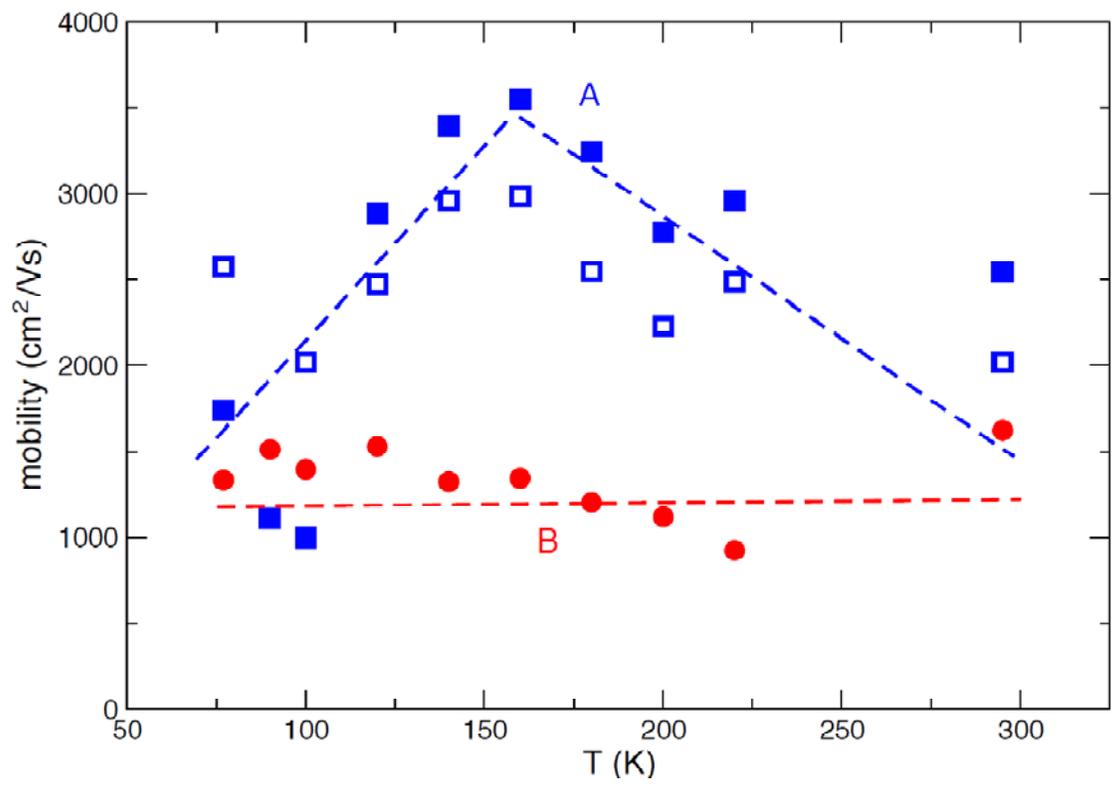

Fig. 2

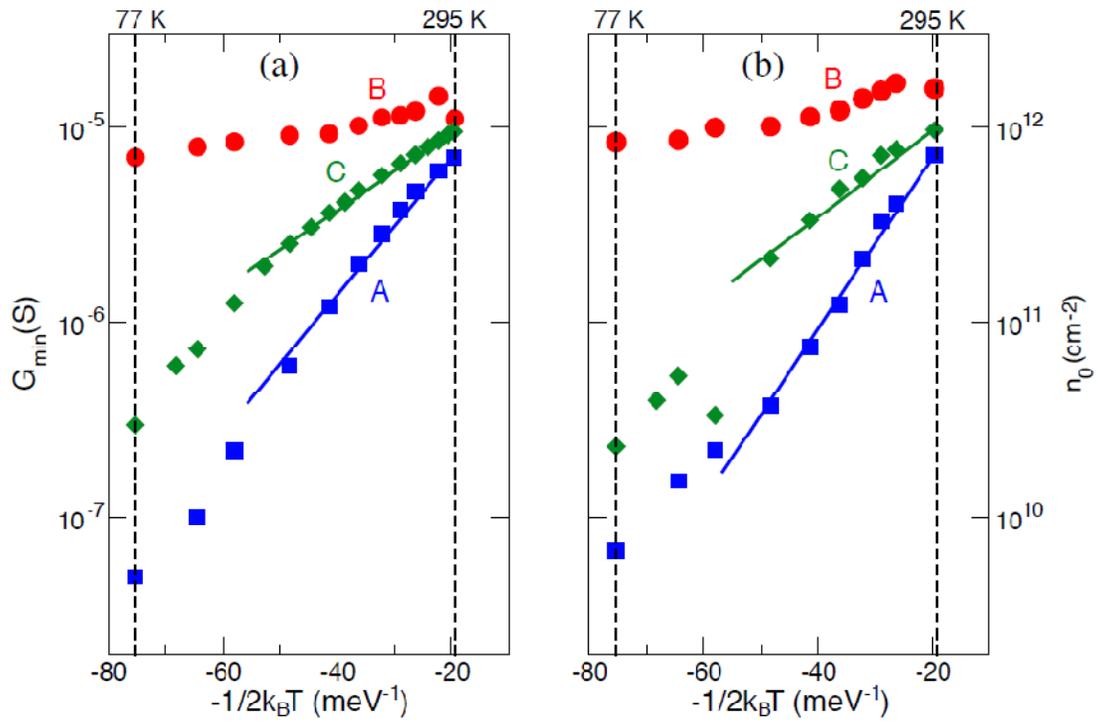

Fig. 3

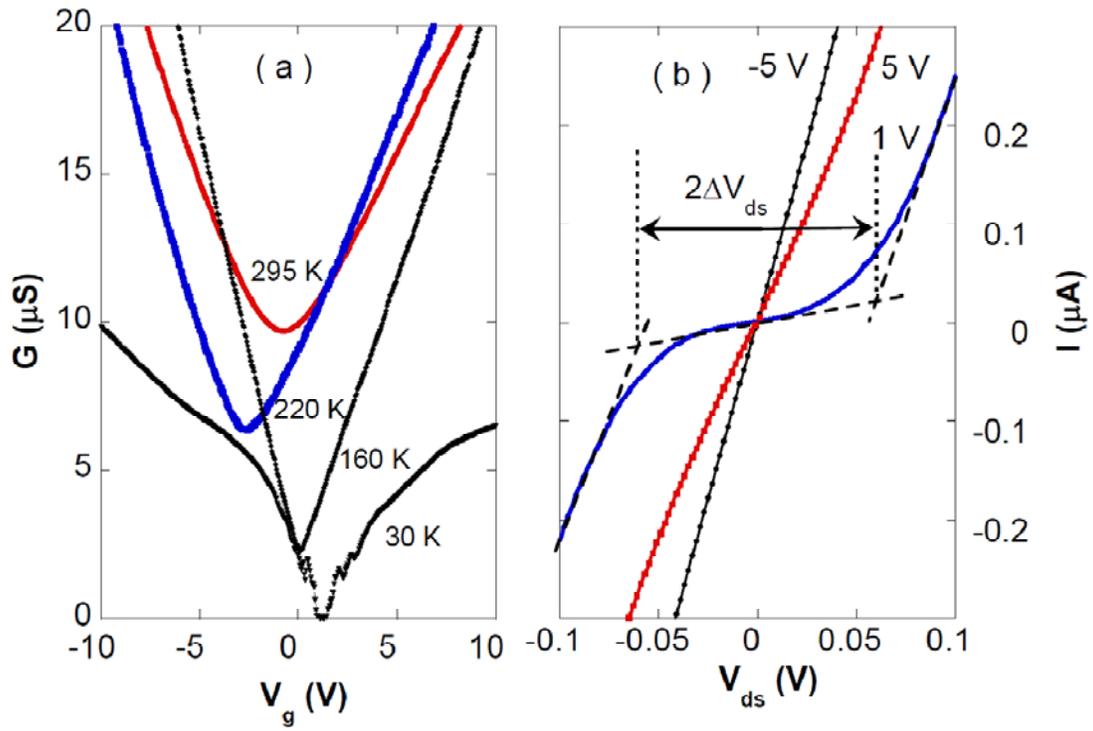

Fig. 4

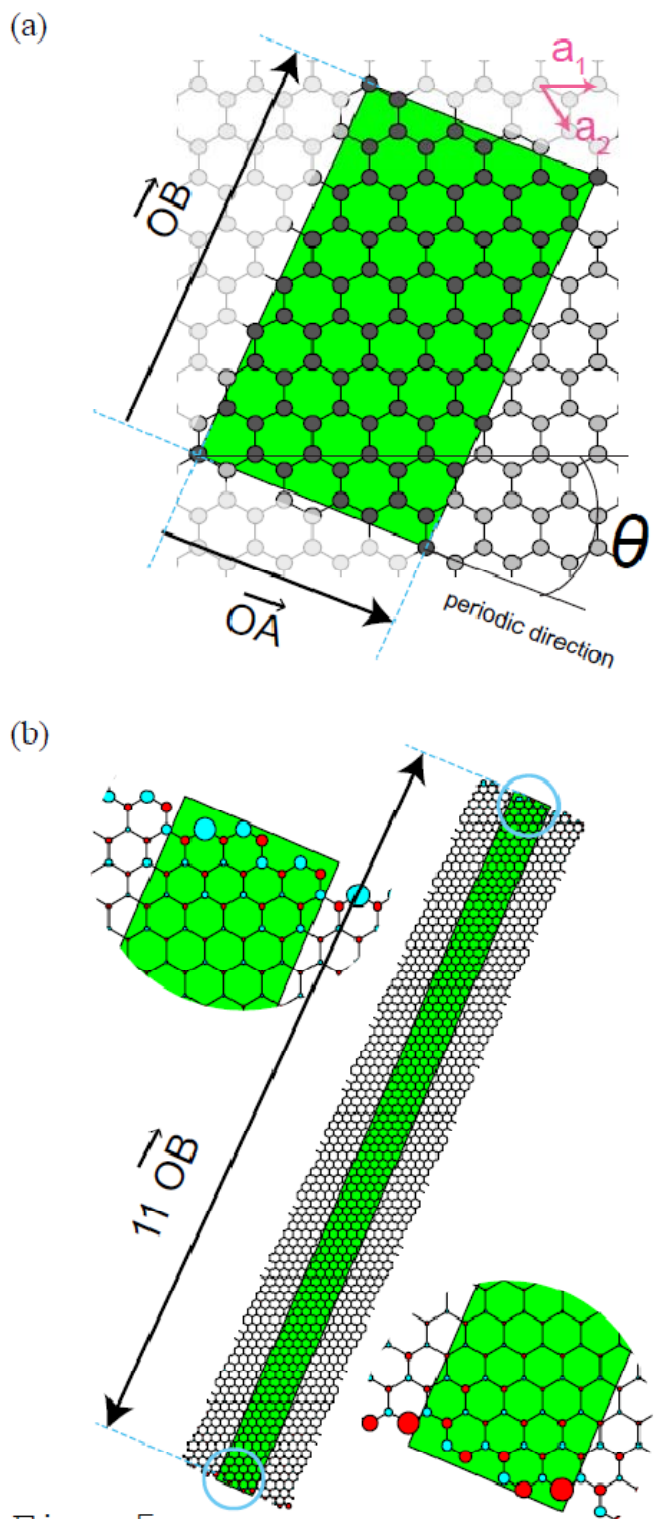

Fig. 5

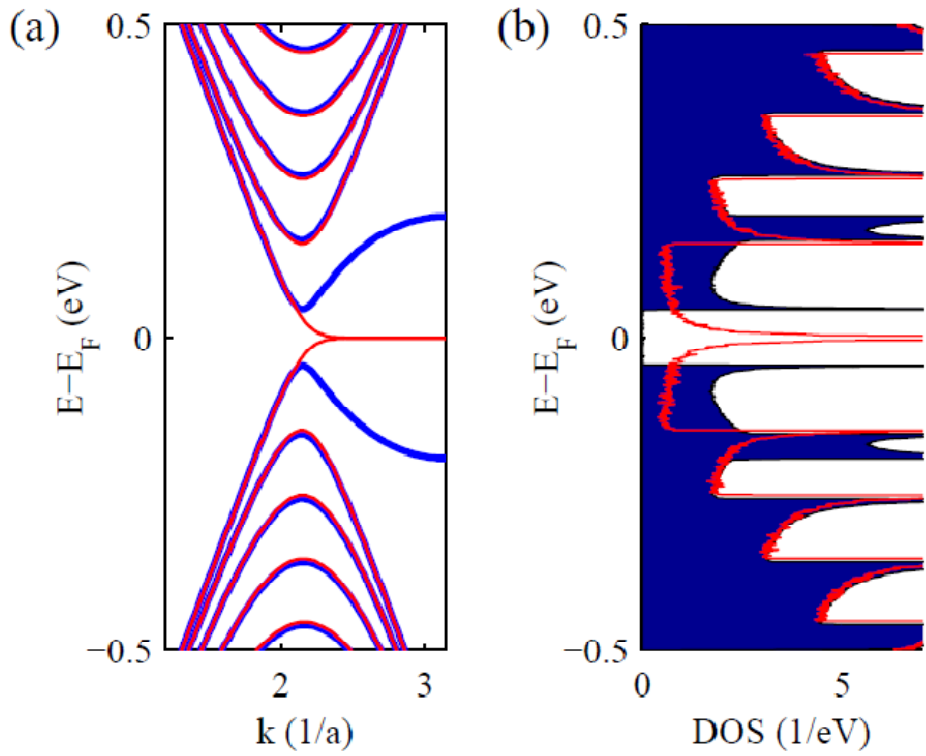
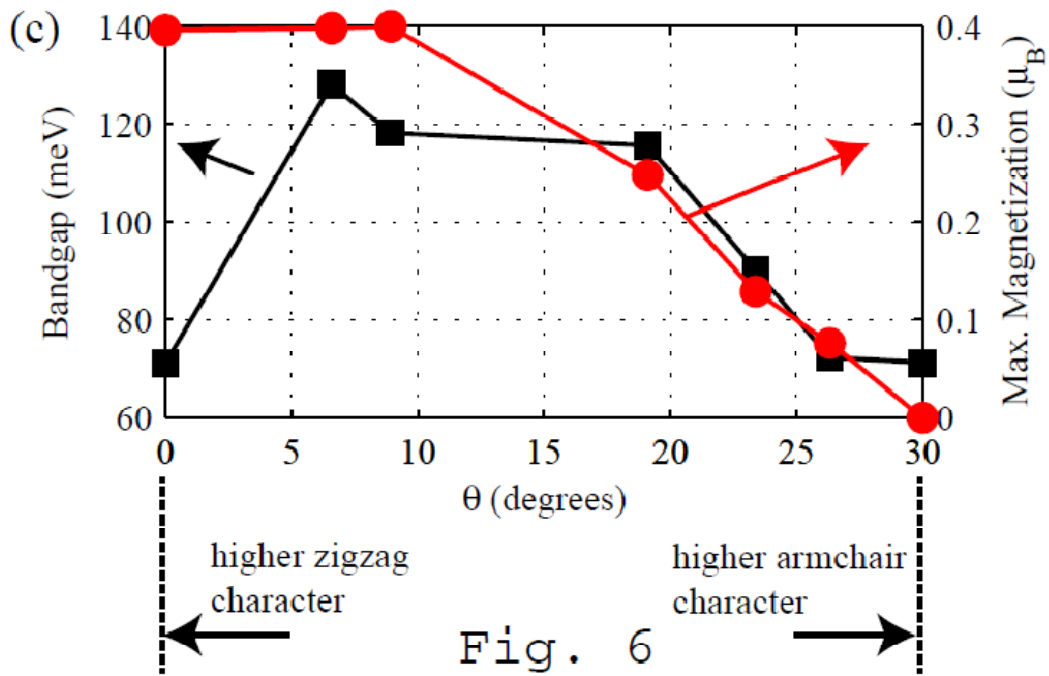

Fig. 6